# Similarity of Precursors in Solid-State Synthesis as Text-Mined from Scientific Literature

Tanjin He[ab], Wenhao Sun[b], Haoyan Huo[ab], Olga Kononova[ab], Ziqin Rong[b], Vahe Tshitoyan[c], Tiago Botari[d], Gerbrand Ceder*[ab]

[a] Department of Materials Science and Engineering, UC Berkeley, Berkeley, CA, 94720, USA
[b] Materials Sciences Division, Lawrence Berkeley National Laboratory, Berkeley, CA, 94720, USA
[c] Google LLC, 1600 Amphitheatre Parkway, Mountain View, CA, 94043, USA
[d] Institute of Mathematics and Computer Sciences, University of São Paulo, São Carlos, SP, 13566-590, Brazil
* Email: gceder@berkeley.edu

**ABSTRACT:** Collecting and analyzing the vast amount of information available in the solid-state chemistry literature may accelerate our understanding of materials synthesis. However, one major problem is the difficulty of identifying which materials from a synthesis paragraph are precursors or are target materials. In this study, we developed a two-step Chemical Named Entity Recognition (CNER) model to identify precursors and targets, based on information from the context around material entities. Using the extracted data, we conducted a meta-analysis to study the similarities and differences between precursors in the context of solid-state synthesis. To quantify precursor similarity, we built a substitution model to calculate the viability of substituting one precursor with another while retaining the target. From a hierarchical clustering of the precursors, we demonstrate that the "chemical similarity" of precursors can be extracted from text data. Quantifying the similarity of precursors helps provide a foundation for suggesting candidate reactants in a predictive synthesis model.

## 1. INTRODUCTION

Understanding how to synthesize the desired compounds is a grand challenge in the development of novel materials[1]. Researchers are trying to tackle this challenge from different perspectives, including *in situ* experiments[2–4], thermodynamic analysis[5–8], and machine learning-guided synthesis parameter search[9,10]. One potential approach is to learn from the large volume of experimental synthesis "recipes", which are provided in scientific publications in various unstructured forms[11–14]. Here, we define a solid-state synthesis recipe to be any structured information about a target material, precursors, and operations used to synthesize this material, that is classified as solid-state synthesis by the decision tree approach of Huo *et al.*[15]. In order to understand and eventually predict solid-state synthesis recipes, one of the important questions is how to select precursors. Knowledge of which precursors to use is often achieved by an individual's experience. Here we present a data-driven approach to assess the similarities and differences between precursors in solid-state synthesis by extracting precursors and targets from literature and conducting a meta-analysis with the extracted data.

The extraction of precursors and targets from written text is difficult because of the complexities of natural language. First, a material entity can be written in text in various complicated forms; they can be represented as chemical formulas such as $Al_2O_3$ and $A_xB_{1-x}C_{2-\delta}$, chemical terms such as hafnium oxide, acronyms such as PZT for $Pb(Zr_{0.5}Ti_{0.5})O_3$, and even more complicated notations for composites and doped materials such as $Si_3N_4$-30 wt% $ZrB_2$ and $Zn_3Ga_2Ge_{2-x}Si_xO_{10}$:2.5mol% $Cr^{3+}$. Translating this knowledge into explicit rules for Chemical Named Entity Recognition (CNER) is difficult.

Second, material entities can play different roles in synthesis experiments such as targets, reagents, reaction media, and so forth. While this can usually be recognized easily by researchers based on their domain-specific knowledge and grammar comprehension, such an implicit assignment of meaning is much harder in computational algorithms. One naïve approach could be to use multiple rules to distinguish between targets and precursors. For example, assign a simple material (e.g., $TiO_2$) as a precursor and a complex material (e.g., $Pb(Zr_{0.5}Ti_{0.5})O_3$) as a target, because researchers usually use simple materials to synthesize a complex one. However, there are many cases that do not follow this rule: the same material zirconia can be a precursor for a Zr-based complex oxide, an auxiliary component as a grinding media, or even a target in the synthesis of stabilized or doped zirconia[16]. In order to correctly identify if a material plays the role of target, precursor, or something else, one needs to read the context of the sentence or entire paragraph, in addition to finding the material expressions. Hardcoding all possible rules would require an enormous amount of human effort.

Recent progress in natural language processing (NLP)[17,18] has made it possible to locate words or phrases in unstructured text and classify them into pre-defined categories. For example, Swain *et al.* trained a conditional random field (CRF) model on an organic dataset[19] to extract chemical entities—available in the toolkit ChemDataExtractor.[20] Kim *et al.* utilized a neural network trained on 20 articles to extract 18 different categories of synthesis information, including materials and targets, for 30 different oxide systems[21]. Korvigo *et al.* developed a CNN-RNN model to extract chemical entities[22] on the same dataset as Swain *et al.*. Weston *et al.* trained a bi-directional long-short term memory (Bi-LSTM) model to extract inorganic materials from materials science abstracts[23]. Other packages to extract chemical entities using NLP methods include OSCAR4[24], ChemicalTagger[25], GRAM-CNN[26], and so forth. However, the previous studies mainly focused on the identification of chemicals rather than their roles in synthesis. Kim *et al.*[21] demonstrate an attempt to predict and analyze targets.

Our focus here is specifically to identify precursor and target materials in inorganic solid-state synthesis text, and to study the relations between various precursors and correlate them with targets. For the CNER task, a two-step model Synthesis Materials Recognizer (SMR) based on Bi-LSTM is implemented. The model recognizes context clues provided by the words around the precursors/targets in the sentence. With the SMR model, we extracted 1,619 unique precursors and 16,215 unique targets from 95,283 paragraphs in 86,554 scientific papers on solid-state synthesis. This corpus of papers was filtered from a larger set of 4 million papers as described in ref Kononova *et al.*[27]

Quantitative analysis of this large-scale dataset indicates that the most common precursors for each element are usually the oxides, carbonates, or hydroxides stable at ambient environment. By applying a probabilistic model on the data we explore which precursors play a similar role in the synthesis of a target material and which may therefore be substitutable. Combining the substitution probability and the distribution of synthesis temperatures, we define a multi-feature distance metric to characterize the similarity of precursors. A hierarchical clustering of precursors based on this metric demonstrates that the "chemical similarity" can be extracted from text data, without the need to include any explicit domain knowledge. The quantitative similarity metric offers a reference to rank precursor candidates and constitutes an important step toward developing a predictive synthesis model.

## 2. EXTRACTION OF PRECURSORS AND TARGETS

In this section, we describe the SMR model used to identify and extract precursor and target materials from a synthesis paragraph. By comparing with a baseline model, we explain how the SMR model works and its advantages and limitations.

**2.1 Data Preparation.** We used the same data extraction pipeline as described in ref Kononova *et al.*[27] A total of 4,061,814 papers were scraped from main publishers including Elsevier, Wiley, Springer, the American Chemical Society, the Electrochemical Society, and the Royal Society of Chemistry. After classification using the semi-supervised random forest model from Huo *et al.*[15], 371,850 paragraphs in the experimental sections were found to describe inorganic synthesis, such as solid-state, hydrothermal, sol-gel, co-precipitation syntheses, with 95,283 of them corresponding to solid-state synthesis. These 95,283 paragraphs and their corresponding abstracts from 86,554 literature papers were used for materials extraction.

**2.2 Algorithm Design and Execution.** The identification of material entities in the text and their subsequent classification as targets, precursors, or something else were performed in two steps, as shown in Figure 1(a): first we identified all material entities present in a sentence; next we replaced each material with a keyword "<MAT>" and classified it as a "Target", "Precursor", or just "Material". Each step was executed by a different Bi-LSTM neural network with a CRF layer on top of it (Bi-LSTM-CRF)[28,29].

For the first step, each input word was represented as the combination of a word-level embedding and a character-level embedding via an embedding layer. The word-level embeddings, which are vectors of real numbers representing the words, were trained using the Word2Vec approach[30,31] with ~33,000 paragraphs on solid-state synthesis to capture the semantic and syntactic similarity of words in synthesis text. In this embedding layer, the characters of each word were converted into an embedding vector using another Bi-LSTM to learn the character-level features such as the prefix and suffix information. The character embedding was concatenated with the pretrained word embedding and input into a Bi-LSTM to capture the left and right context at every word. Finally, the output from Bi-LSTM was combined with a CRF model, which characterized the transition probability from one tag to another to produce the final prediction.

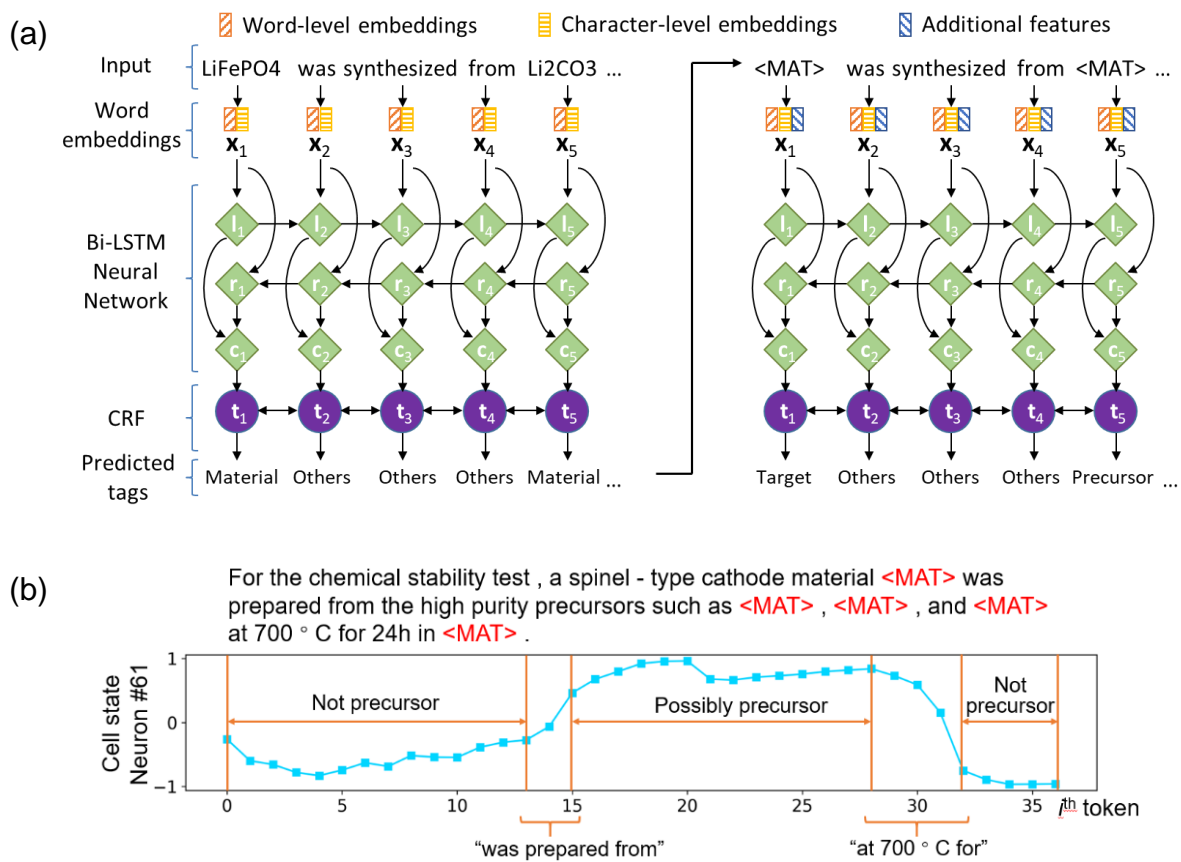

**Figure 1.** (a) Main architecture of the SMR model. $x_i$ is the embedding used as the input for the Bi-LSTM-CRF neural network. $l_i$ represents the $i^{th}$ token and its left context. $r_i$ represents the $i^{th}$ token and its right context. $c_i$ is the combination of $l_i$ and $r_i$. $t_i$ represents the score for different tags. (b) Change of one LSTM cell state in different context for precursor classification. The tokens in the example sentence are separated by spaces in the hanging text and represented as the sequence numbers on the x-axis.

For the second step, a Bi-LSTM with a similar structure to that in the first step was used but the inputs were different. All the materials in the input sentences were replaced with the word "<MAT>" so that the role of a material in synthesis is predicted mainly based on the surrounding context. We found this to be more effective than directly using the specific materials words as input to the Bi-LSTM, because such a direct model tries to store the mapping information from each different material to the particular role this material is mostly used for, which brings in bias for frequently appearing materials. For example, as "zirconia" often describes the balls used in ball milling, it is difficult for the neural network to deviate from this assignment and treat "zirconia" as a target or precursor. A more detailed discussion on the benefits of the two-step model can be found in the SI. Since all the chemical information about the material is lost by inputting "<MAT>" instead of the materials words, we also included two additional features in the word representation, that is, the number of metal/metalloid elements and a flag indicating whether the material contained C, H, and O elements only. These additional features assist in the differentiation of precursors and targets, as they tend to have different numbers of metal/metalloid elements and are generally not organic compounds in inorganic synthesis. The composition information was obtained by parsing the raw text of the material entities by regular expression comparison[27].

Bi-LSTM is able to infer the role of materials from context because Bi-LSTM specifies a variable called cell state to store the information about the words around the material. Figure 1(b) shows a typical example of the trained Bi-LSTM cell state continuously changing depending on token context in the example sentence[32] when feeding the tokens into Bi-LSTM one by one. In this study, 100 neurons (cells) were used to represent the context information; Figure 1(b) displays one of the cell states relevant to the context of precursors. To obtain the cell states for the next token, both the next token and the current cell states are input to the network. Hence, after seeing the sequence of tokens "was prepared from" in the example sentence, the network predicts from the context that the tokens following this phrase most likely refer to a precursor(s). Likewise, the network predicts that the tokens following "at 700 ° C for" most likely are not precursors.

To train the SMR model, 834 solid-state synthesis paragraphs from 750 papers were tokenized with ChemDataExtractor[20], and each token was manually annotated with tags of "Material", "Target", "Precursor", and "Outside" (not a material entity). In the annotation, a target is defined as a final material obtained through a series of lab operations in the complete synthesis process, and a precursor is defined as a starting reagent involved in the synthesis process through a lab operation and contributing to the target composition. Other materials include media, gas, device materials, and so forth. The annotation dataset contains 8,601 materials, out of which 1,256 are targets and 3,295 are precursors. The annotated dataset was randomly split into training/validation/test sets with 500/100/150 papers in each set. Early stopping[33] was used to minimize overfitting by stopping the iterative training when the best performance was achieved on the validation set. To reduce the variance resulting from the limited size of the training set, the six models trained in a six-fold cross-validation process were combined together to make the final decision by voting in the classification. The entire training and test process was repeated 10 times, and the average result of the test sets is reported.

**2.3 Evaluation of SMR Accuracy and Working Examples**. We first aim to demonstrate that the recognition of context clues is necessary for the CNER task by comparing the SMR model with a baseline model based on naïve rules. To build this baseline model, we used ChemDataExtractor[20] to identify and extract materials from the text. Then, inspired from a scientific perspective that researchers usually use simple materials to synthesize a complex one, the precursors and targets were assigned based on the number of elements: materials with only one metal/metalloid element were assumed to be precursors, and materials with at least two metal/metalloid elements were assumed to be targets. This baseline is a least-effort model but provides a quantitative reference for understanding the importance of capturing context information.

In Table 1, we compare the performance of the SMR model and the baseline model using $F_1$ scores, which provides a measure of the accuracy of a binary classification test based on the harmonic mean of the precision and recall. The $F_1$ scores on the extraction of all materials, precursors, and targets using the SMR model are 95.0%, 90.0%, and 84.5%, respectively. Out of all the extracted entities, 88.9% of precursors and 85.9% of targets in the test set are correctly identified. These correct cases account for 91.2% and 83.4% of all the precursors and targets which should be extracted, respectively. The possibility of errors increases when multiple precursors and targets are present in the same sentence. Out of all the sentences containing precursors/targets, the rate to successfully retrieve *all* the precursors and targets in each sentence is 73.4%. Some representative successful examples from the SMR model, such as the recognition of the targets "LiBaBO3:Sm3+" and "(0.725-x)BiFeO3-xBi(Ni0.5Mn0.5)O3-0.275BaTiO3 + 1 mol% MnO2", are shown in Table 2.

We interpret the results as follows. In the baseline model, only the information from the material entity itself is used, resulting in low $F_1$ scores for the extraction of precursors and targets (70.0% and 32.1%, respectively). In contrast, the SMR model achieves better $F_1$ scores because Bi-LSTM is able to infer the role of materials from the context. For example, as discussed previously, the Bi-LSTM infers from the tokens "was prepared from" to mean that the following tokens probably refer to a precursor(s). Likewise, the network predicts that the tokens following "at 700 ° C for" most likely are not precursors. For a precursor with more than one metal/metalloid element, the baseline model fails to recognize it regardless of the context, while the SMR model can still identify the precursor nature of this material.

**Table 1.** Precision, recall, and $F_1$ scores for the baseline and SMR models to extract materials, precursors, and targets. The type "Materials" include precursors, targets, and all other materials.

| Model | Type | Precision (%) | Recall (%) | $F_1$ score (%) |
| --- | --- | --- | --- | --- |
| Baseline | Materials | 78.3 | 68.3 | 73.0 |
| | Precursors | 60.9 | 82.2 | 70.0 |
| | Targets | 48.5 | 33.0 | 32.1 |
| SMR | Materials | 94.6 | 95.3 | 95.0 |
| | Precursors | 88.9 | 91.2 | 90.0 |
| | Targets | 85.9 | 83.4 | 84.5 |

However, some situations still remain difficult for the SMR model:

• Some material entities tokenized into multiple tokens are not completely extracted. For example, the incomplete pieces "(Ba1-x(K" and "Na)x/2Lax/2)(Mg1/3Nb2/3)O3" are extracted instead of "(Ba1-x(K or Na)x/2Lax/2)(Mg1/3Nb2/3)O3", as listed in Table 2. The identification of these materials is difficult because of the syntactic variability and ambiguity of multiword expressions (MWEs)[34], which might be improved by incorporating recent progress on MWE identification such as the language-independent architecture proposed by Taslimipoor *et al*.[35] The number of training sentences containing MWE materials might remain as an issue considering the relatively large dataset[36] used by Taslimipoor *et al*.[35]

• Some sentences are ambiguous to the SMR model because of the limitations of the training set. For example, the model correctly classifies "Y2O3" as a precursor in "Y2O3 as a precursor was added" and "Y2O3" as neither target nor precursor in "Y2O3 as a grinding media was added". However, in the sentence "Y2O3 as a donor impurity was added", the model does not understand "donor impurity" and only assigns "Y2O3" as an ordinary material rather than a precursor. This situation might be improved by including more contextual information in the input, such as the sentence embeddings[37] of previous and next sentences, and contextualized word embeddings trained on a much larger corpus (e.g. BERT[38] and SciBERT[39]). Future possible directions for research include training these embedding models on papers specifically on materials synthesis, although the training process may require a significant manual time investment and considerable computational resources.

• Misclassification can occur when the sentence is written with a complicated structure. For example, the target "Ba0.5Sr0.5CoxFe1-xO3-δ" is misclassified as a precursor when the order of precursors and targets is reversed or closely mixed in the sentence and the materials around this word are all precursors, as shown in Table 2. These sentences with a complicated structure must often be treated on a case-by-case basis, and it is difficult for an NLP model to pick up general rules to correct for these errors. A potential solution is to conduct selective sampling to annotate sentences with complex syntax more efficiently, where only the ones that a pretrained classifier is less confident with will be sampled for annotation[40]. Our current model lays a foundation for selective sampling.

Considering the significant effort required to address each of these problems and the decent performance achieved already, we put these problems as future research directions. To retain a higher precision in the dataset, we only used the recipes for which a balanced chemical reaction can be reconstructed from the extracted precursors and targets as described by Kononova *et al.*[27]

**Table 2.** Representative successful and failed examples from the SMR model in this study.

| Example Sentences | Expected | Error in Extraction |
|---|---|---|
| *Successful* | | |
| The LiBaBO3:Sm3+ samples were prepared by solid-state reaction.[41] | Target: LiBaBO3:Sm3+ | N/A |
| Ceramic samples of (0.725-x)BiFeO3-xBi(Ni0.5Mn0.5)O3-0.275BaTiO3 + 1 mol% MnO2 (x = 0-0.08) (BFO-BT-BNM-x) were prepared by the conventional solid-state route using high-purity metal oxides and carbonates as starting materials: Bi2O3 (99 %), Fe2O3 (99 %), BaCO3 (99 %), TiO2 (98 %), NiO (99 %), MnO2 (99.99 %).[42] | Targets: (0.725-x)BiFeO3-xBi(Ni0.5Mn0.5)O3-0.275BaTiO3 + 1 mol% MnO2, BFO-BT-BNM-x<br>Precursors: Bi2O3, Fe2O3, BaCO3, TiO2, NiO, MnO2 | N/A |
| Y2O3 as a precursor was added. | Precursor: Y2O3 | N/A |
| Y2O3 as a grinding media was added. | Material: Y2O3 | N/A |
| *Failed* | | |
| (Ba1-x(K or Na)x/2Lax/2)(Mg1/3Nb2/3)O3 with 0 ≤ x ≤ 1 were synthesized by a conventional solid-state reaction method.[43] | Target: (Ba1-x(K or Na)x/2Lax/2)(Mg1/3Nb2/3)O3 | "(Ba1-x(K" and "Na)x/2Lax/2)(Mg1/3Nb2/3)O3" extracted |
| Y2O3 as a donor impurity was added.[44] | Precursor: Y2O3 | Y2O3 extracted as an ordinary material |
| Required amounts of BaCO3, SrCO3, CoCO3·0.5H2O and Fe2O3 powders for Ba0.5Sr0.5CoxFe1-xO3-δ, Pr6O11, BaCO3, and CoCO3·0.5H2O powders for PrBaCo2O5+δ were mixed and ball-milled for 24h.[45] | Targets: Ba0.5Sr0.5CoxFe1-xO3-δ, PrBaCo2O5+δ<br>Precursors: BaCO3, SrCO3, CoCO3·0.5H2O, Fe2O3, Pr6O11, BaCO3, CoCO3·0.5H2O | Ba0.5Sr0.5CoxFe1-xO3-δ extracted as a precursor |

## 3. SUBSTITUTION OF PRECURSORS

We first present the variety of the extracted precursors. An intriguing question is how frequently researchers substitute one precursor with another while retaining the target, which sheds light on how similarly these precursors behave in a solid-state reaction. We utilize a substitution model based on the work of Hautier *et al.*[46] and Yang *et al.*[47] to quantify the probability that two precursors are interchangeable.

**3.1 Common and uncommon precursors**. The SMR model was applied to generate the dataset of 29,308 reactions by analyzing 95,283 solid-state synthesis paragraphs (see the work by Kononova *et al.*[27] for details). Since a reaction can be mentioned multiple times in the same paper, resulting in multiple records in the dataset, the records were unified to 28,530 reactions, containing 71 different metal/metalloid elements and 1,619 distinct precursors. Some precursors are rarely used. Restricting the statistics to precursors used at least 30 times, there are 58 metal/metalloid elements and 182 precursors.

To visualize the variety of precursors, the precursors for each metal/metalloid element are categorized by the anion (group) class and counted by the number of corresponding reactions in which they are used. The frequency of each anion class normalized by the total number of reactions for an element is shown in Figure 2. One precursor is usually used much more frequently than other precursors for the same element, which we denote as the common precursor. Figure 2 shows that for alkali and alkaline earth elements, the common precursors are carbonates, except for MgO which is the typical source for Mg. For transition metals and other main group elements, the common precursors are oxides except for $B(OH)_3$ for B. In general, the common precursor tends to be the compound that is stable under ambient conditions, which is beneficial to the purity and accurate weighting in experiments[48]. Our observation on the common precursors suggests that laboratory chemists will prioritize shelf stability of precursors; although we note that more reactive precursors can help to facilitate synthesis reactions.

Sometimes, the decision to use an uncommon precursor is motivated by an interesting advantage for a specific nontraditional precursor. For example, in some cases precursors can function as morphology templates; Zhao *et al.* reported that γ-MnOOH nanorods were used to obtain $LiMn_2O_4$ nanorods, whereas $LiMn_2O_4$ from electrolytic $MnO_2$ (EMD) only consisted of many irregular and aggregated particles[49]. The use of a lower-melting-point precursor can result in a target with a smaller particle size; Liu *et al.* adopted $Sr(NO_3)_2$ instead of $SrCO_3$ to synthesize $SrTiO_3$ nanocrystals[50]. An amorphous precursor can facilitate the reaction process and minimize the possibility of forming chemical segregations; Mercury *et al.* utilized amorphous $Al(OH)_3$ rather than $Al_2O_3$ in the synthesis of $Ca_3Al_2O_6$[51]. In these examples, there were strategically designed precursors in order to achieve a particular synthesis result. Collecting these individual-use cases provides interesting insights into synthesis design.

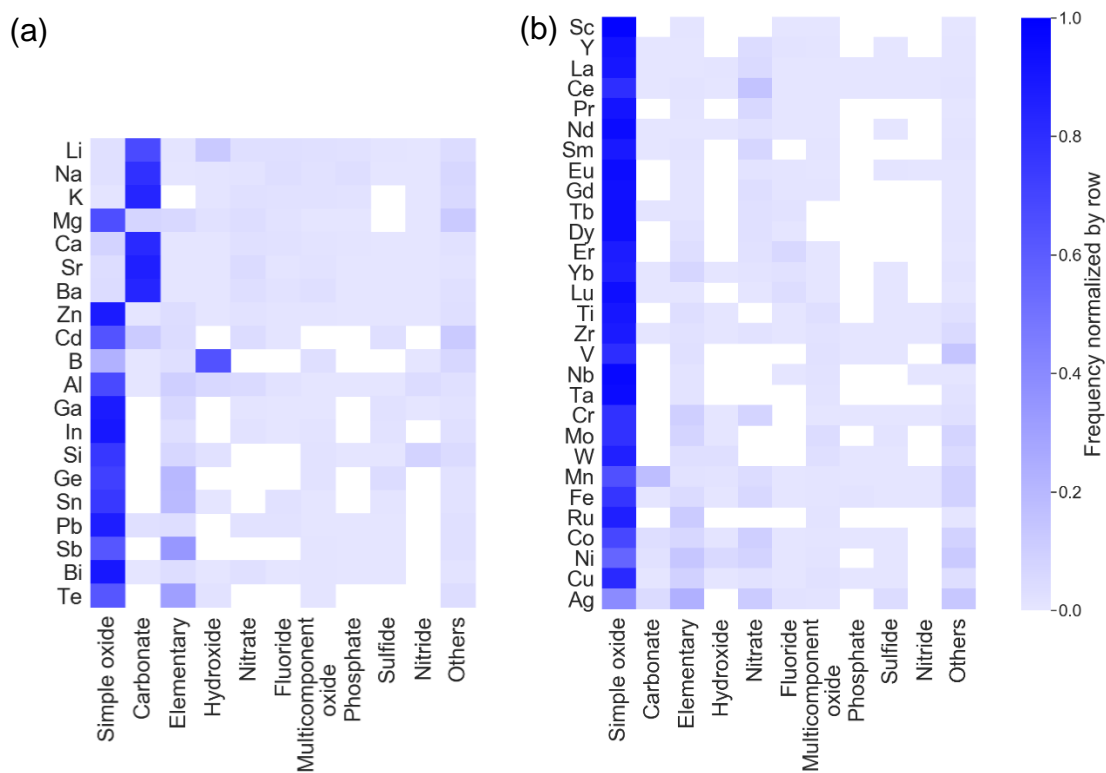

**Figure 2**. Fraction of different classes of precursors corresponding to each element: (a) main group elements and (b) transition metal elements.

**3.2 Substitution model.** The large number of reactions we obtained gives us the opportunity to understand to what extent precursors are interchangeable. To measure the probability that one precursor can be substituted by another while retaining the target, we utilized a substitution model similar to the one developed by Hautier *et al.*[46] and used by Yang *et al.*[47] for structure prediction. For each pair of precursors, the model counts the number of occurrences where the same targets can be synthesized from either of the precursors. The more frequently the two precursors are interchanged, the more similar they are.

In the following part, we define the substitution model in a mathematical form, and express the probability of finding a substitutional precursor pair $P_{\text{sub}}(p_i^{j,1}, p_i^{j,2})$ as a sigmoid with unknown parameter $\lambda$. Assuming the independence of substitutions, we deconvolute the probability of finding substitution between two lists of precursors $P_{\text{sub}}(R_X, R'_X)$ into the product of $P_{\text{sub}}(p_i^{j,1}, p_i^{j,2})$. At last, we maximize $P_{\text{sub}}(R_X, R'_X)$ over substitution observations to solve $\lambda$ and use it to calculate substitution probability.

First, we define precursor substitution in a mathematical form. Let $E=(e_1, e_2, ..., e_n)$ be a pre-defined ordered list of all the metal/metalloid elements given in the periodic table. We assume each precursor contributes one metal/metalloid element to targets. For the target $R_{Tar}$ in a reaction synthesis $R=(R_{Tar}, R_X)$, define the precursor list as $R_X=(p_1, p_2, ..., p_n)$, where $p_i$ is the precursor for element $e_i$ present in $R_{Tar}$; otherwise $p_i$ is null. For a pair of reaction $\{R, R'\}$, if $R_{Tar}=R'_{Tar}$ and $R_X \neq R'_X$, we say precursor substitution occurs. Through iterating over all the possible combinations of any two reactions, we obtain a collection of $N$ reaction pairs where precursor substitution occurs, denoted as the data $D=\{\{R, R'\}^1, \{R, R'\}^2, ..., \{R, R'\}^N\}$. Our objective is now to find the values of the pairwise precursor substitutions that maximize the likelihood of $D$.

Next, we define the potential substitutional precursor pairs. For element $e_i$, denote the list of candidate precursors as $(p_i^1, p_i^2, ..., p_i^{m_i})$, where $m_i$ is the total number of unique precursors. We assume that potentially every precursor $p_i^{\tau_1}$ can be substituted by any other one $p_i^{\tau_2}$, forming a substitutional pair $\{p_i^{\tau_1}, p_i^{\tau_2}\}$ where $1 \leq \tau_1 < \tau_2 \leq m_i$. In total, there can be up to $M_i = \binom{m_i}{2}$ such pairs for element $e_i$. For simplicity, we assemble all substitutional pairs for all elements into one list and renumber the pairs as $\{p_i^{j,1}, p_i^{j,2}\}$ where $j = 1, ..., \sum_{i=1}^{n}|M_i|$. Although the index $i$ is not necessary, we retain it for clarity to distinguish between elements. The probability that the pair $\{p_i^{j,1}, p_i^{j,2}\}$ can be found as a substitution occurs is written as

$$P_{\text{sub}}(p_i^{j,1}, p_i^{j,2}) = \text{sigmoid}(\lambda_j) \tag{1}$$

where $\lambda_j$ is a parameter to be optimized. Assuming all substitutional precursor pairs are independent of each other, the probability that the pair of precursor lists $\{R_X, R'_X\}$ can be found as a substitution occurs is

$$P_{\text{sub}}(R_X, R'_X) = \frac{e^{\sum_j \lambda_j \mathbf{1}_j(R_X, R'_X)}}{Z} \tag{2}$$

where

$$\mathbf{1}_j(R_X, R'_X) = \begin{cases} 1, \{R_{X,i}, R'_{X,i}\} = \{p_i^{j,1}, p_i^{j,2}\} \\ 0, otherwise \end{cases} \tag{3}$$

and $Z$ is the partition function for normalization, given by

$$Z = \prod_j (1 + e^{\lambda_j}) \tag{4}$$

The value of $\boldsymbol{\lambda} = (\lambda_1, \lambda_2, ...)$ is obtained by maximizing the likelihood over the data $D$:

$$\boldsymbol{\lambda}^* = \text{argmax}_\lambda \sum_{t=1}^{N} \log P_{\text{sub}}((R_X, R'_X)^t | \lambda) \tag{5}$$

For those substitutional pairs not found in $D$, the value of $\lambda_j$ will be set to a common low value such that $P_{\text{sub}}(p_i^{j,1}, p_i^{j,2})$ in Eq. (1) is close to zero.

Finally, we define the substitution probability. Here we discuss one substitutional pair $\{p_i^{j,1}, p_i^{j,2}\}$ and omit the index $j$ for simplicity. For a given reaction using precursor $p_i^1$, the probability that $p_i^1$ is substitutable by $p_i^2$ is

$$P(p_i^2|p_i^1) = P(p_i^1 \text{ substituted}) \frac{P_{\text{sub}}(p_i^1, p_i^2)}{\sum_{k \neq 1} P_{\text{sub}}(p_i^1, p_i^k)} \tag{6}$$

where $P(p_i^1$ substituted$)$ is a prior probability of $p_i^1$ being substitutable and is calculated as the number of reactions with the substituted precursor $p_i^1$ divided by the total number of all reactions using $p_i^1$. The fractional part in the right-hand side accounts for the conditional probability that $p_i^1$ is substitutable by $p_i^2$ when substitution occurs, which can be calculated with Eq. (1). A small fraction of reactions (~5%) which included multiple metal/metalloid elements in the same precursors or used multiple precursors for the same element were not considered in this model.

**3.3 Cross-validation of the substitution model.** We evaluated the predictive power of the substitution model by performing a cross-validation test on the generation of alternative precursor lists. Cross-validation consists in training the model on part of the available data (the training set) and predicting back the remaining data (the validation set). Given a target $R_{Tar}$ and an existing precursor list $R_X$ in the training set, we can propose an alternative precursor list $R'_X$ to synthesis the same target by replacing the precursors in $R_X$ with different ones. With the substitution probability defined in Eq. (6), the conditional probability of $R_X$ being substitutable by $R'_X$ is given by

$$P(R'_X|R_X) = \prod_{p_i^1 \in R_X, p_i^2 \in R'_X, p_i^1 \neq p_i^2} P(p_i^2|p_i^1) \tag{7}$$

If $P(R'_X|R_X)$ is higher than a given threshold, the proposed $R'_X$ will be accepted as a positive prediction of an alternative precursor list. Otherwise, $R'_X$ will be rejected as a negative prediction. Applying this procedure on all possible $R'_X$, we obtain all the positive and negative predictions and compare with the validation set for evaluation. Two-thirds of the reactions were used as the training set and the remaining one-third of the data were used as the validation set. For example, $La_{0.7}Ca_{0.3}MnO_3$ is synthesized from $La_2O_3$, $CaCO_3$, and $MnO_2$[52] in the training set. As a true positive prediction, the substituted precursor list $La_2O_3$, CaO, and $Mn(Ac)_2$[53] was also found in the validation set. The true positive rate (TPR) and false positive rate (FPR) were used as metrics to evaluate the performance. The TPR and FPR of the prediction vary with the probability threshold, as shown in Figure 3. Overall, the TPR is higher than the FPR, indicating that the substitution model has a predictive power in the selection of alternative precursors and can effectively distinguish between the substitutions leading to existing precursor lists and those leading to nonexistent ones. Higher threshold values lead to fewer false alarms but imply fewer true hits. An adequate threshold can be found by selecting the one resulting in relatively higher TPR and lower FPR.

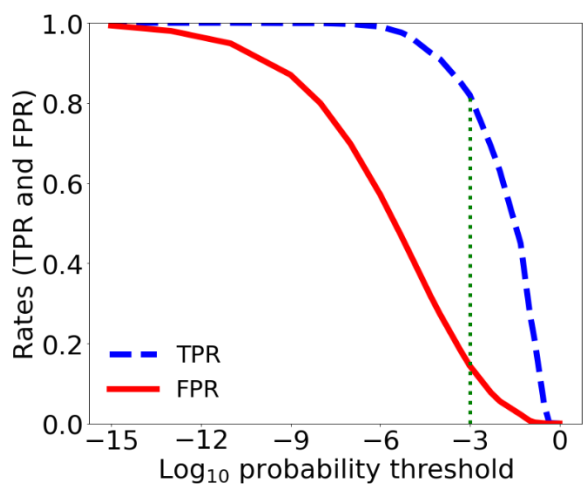

**Figure 3.** TPR and FPR with varying probability threshold in the prediction of alternative precursor list. The green dashed line indicates where the largest difference between the TPR and FPR was observed.

**3.4 Substitution probability.** The probability *P(B|A)* that a precursor A is substituted by another precursor B for the same metal/metalloid element is displayed as a heatmap in Figure 4, where the

rows are A and the columns are B. The color represents the probability of substitution defined in Eq. (6), as shown by the colorbar. For each element, the precursors are ordered by the number of reactions using it from the most to the least, that is, the first precursor is the common precursor for each element. For the sake of simplicity, we merged the precursor in its hydrated form and its anhydrous form, for example LiOH·$H_2$O and LiOH, based on the assumption that water will evaporate early on during the solid-state heating process. The rows for the common precursors usually display relatively high substitution probability, which implies that many uncommon precursors can be replaced with the common precursors. Note that our analysis only indicates that substitution can lead to the same target compound under similar reaction conditions. The choice of different precursors can still be justified as they might infer different properties on the compound. For example, in the battery chemistry, LiOH is sometimes preferred over $Li_2CO_3$ as it leaves less carbonate residual on the surface of the particles.

Intuitively, hydroxides are similar to oxides; however, Figure 4 also captures some differences in this similarity for different elements. For example, the common precursor for Al is the oxide, whereas that for B is the hydroxide. Furthermore, the probability of substitution between $Al(OH)_3$ and $Al_2O_3$ is considerably higher than between $B(OH)_3$ and $B_2O_3$. The number of reactions using $Al_2O_3$, $Al(OH)_3$, $B(OH)_3$, and $B_2O_3$ are 1,606, 148, 705, and 252, respectively, indicating that this difference is not due to limited data. The reason behind this is possibly correlated with the unique bonding in $B_2O_3$; B is highly hybridized with O in $B_2O_3$, much more than Al with O in $Al_2O_3$. This creates strong units in $B_2O_3$ held together by relatively weak forces[54] accounting for its low melting point and high glass-forming ability[55]. Although nitrates are often used in solution-based synthesis, the chance to use nitrates in solid-state synthesis is also considerable. Figure 4 shows that for elements Ca, Ba, Al, and Fe, nitrates frequently replace the common oxide or carbonate precursors. For example, the probability of substituting $Fe_2O_3$ with $Fe(NO_3)_3$ is high. The nitrates are used in various ways such as in conventional solid-state synthesis[56], modified solid-state synthesis[57], combustion synthesis[58], and sol-gel synthesis[59]. Although carbonates appear interchangeable with oxides, the metals in them might not occupy the same valence state. The probability of substitution between $MnCO_3$ and $MnO_2$ is higher than that between $MnCO_3$ and $MnO$, indicating that $MnO_2$ is more similar to $MnCO_3$ than $MnO$.

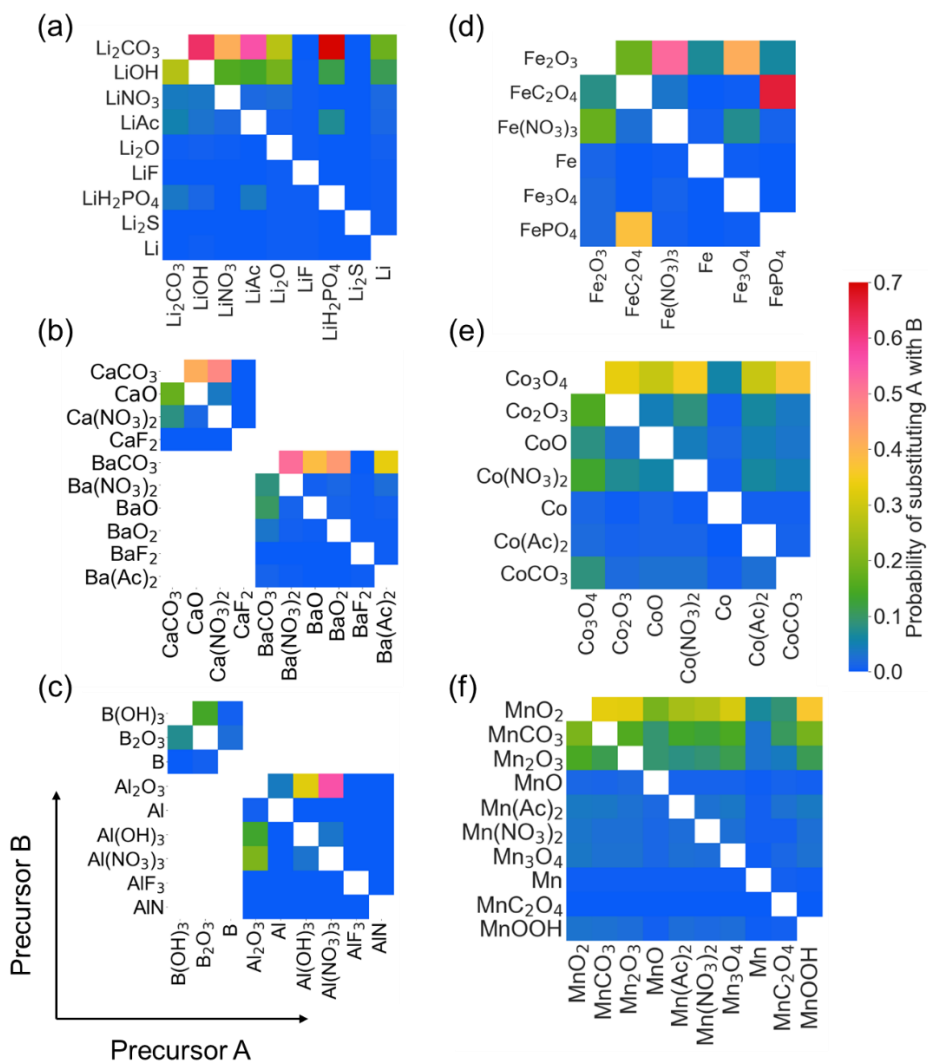

**Figure 4.** Substitution probability $P(B|A)$, which is the probability that the precursor A on the x-axis is substituted with precursor B on the y-axis: (a) Li, (b) Ca and Ba, (c) B and Al, (d) Fe, (e) Co, (f) Mn. For example, we found that in 15% of reactions that use $CaCO_3$, it could also be substituted with another precursor to introduce Ca into the same targets; in 73% of the substitutions, the other precursor is CaO. The joint probability that $CaCO_3$ is substituted and the substitute is CaO is 11%. Because $CaF_2$ is exclusively used for the synthesis of fluorine-containing compounds, the probability that $CaF_2$ is substituted to synthesize the same target is zero.

To better understand how precursors are chosen for elements with variable valence, for each Mn precursor with reasonable frequency of use, we plot in Figure 5 the distribution of valence states for Mn in the targets synthesized from that precursor. The valence of Mn in the target compound was determined by iterating all possible combinations of valence states and finding the one resulting in the charge neutrality for the compound[60]. The width of each violin plot is proportional to the probability density for different valence states; the total area is proportional to the number of reactions using the corresponding precursor. The adoption of MnO, $Mn_2O_3$, and $MnO_2$ is preferred in the literature to synthesize targets with similar valence states, that is, most Mn ions in targets from MnO, $Mn_2O_3$, and $MnO_2$ correspond to 2+, 3+, and 3+~4+, respectively. Different from the oxides, the valence states in targets from $MnCO_3$ and $Mn(Ac)_2$ (Ac stands for acetate anion $CH_3COO^-$) are more evenly distributed, indicating that the use of $MnCO_3$ and $Mn(Ac)_2$ is less dependent on the valence states in the targets. This appears reasonable given the ease by which $MnCO_3$ and $Mn(Ac)_2$ decompose when heated and $Mn^{2+}$ can be oxidized to whatever is stable in the high-component solid under proper oxygen chemical potential. This observation is consistent with the higher probability of substitution between $MnCO_3$ and $MnO_2$ as aforementioned. By comparing the number of reactions using different precursors, it should be noted that the most frequently used Mn precursor to synthesize targets with Mn valence states lower than 3+ remains $MnO_2$, which is the common precursor for Mn, even though $MnO_2$ is more frequently used to synthesize targets with Mn valence states between 3+ and 4+. One possible reason is that Mn at high temperature can rapidly reduce or oxidize driven by the extent of entropic stabilization of $O_2$ on the right-hand side of the reaction $MnO_2 + \Delta H \rightleftharpoons MnO_{2-x} + \frac{x}{2}O_2$. In other words, the metal valence state in the precursor does not necessarily impose the valence state in the target in solid-state synthesis.

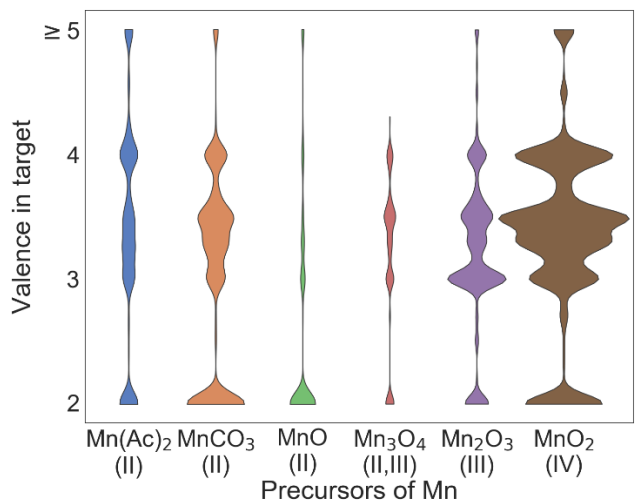

**Figure 5.** Mn valence states in targets from Mn(Ac)$_2$ (manganese acetate), MnCO$_3$, MnO, Mn$_3$O$_4$, Mn$_2$O$_3$, and MnO$_2$. The width in each violin plot is proportional to the probability density for valence at different values. The total area of each violin plot is proportional to the number of reactions using the corresponding precursor.

## 4. SIMILARITY OF PRECURSORS

While *substitutionability*, discussed in the previous section, indicates that a solid-state reaction to the target is possible with the substitutional precursors, it makes no statement as to whether the reaction condition needs to be modified. In the following section we define the *similarity* of precursors based on the *substitutionability* as well as the extent to which the reaction conditions are similar. At this point, we only use temperature to describe the reaction condition considering the amount of effort, but one could extend this concept to capture other synthesis info such as atmosphere, time, number of operations, milling speed, and so forth.

**4.1 Metric for similarity.** Two features, the substitution probability and the distribution of synthesis temperatures of the reactions that use a particular precursor, were utilized to characterize the *similarity* of precursors.

As introduced in Section 3, a precursor $p_i^1$ is substituted by another precursor $p_i^2$ with the probability $P(p_i^2|p_i^1)$. We use the geometric average of $P(p_i^2|p_i^1)$ and $P(p_i^1|p_i^2)$ to balance the asymmetric situations where $p_i^1$ or $p_i^2$ is substituted. The distance accounting for the substitution probability is defined as

$$d_{\text{sub}}(p_i^1, p_i^2) = 1 - \sqrt{P(p_i^1|p_i^2)P(p_i^2|p_i^1)} \tag{8}$$

where $p_i^1$ and $p_i^2$ are two precursors for element $e_i$.

A different precursor can be used with a different synthesis temperature. As an example, the distribution of the highest firing temperature used in synthesis reactions with two different Fe or Ca precursors is presented in Figure 6. The temperatures were extracted by regular expression matching in the corresponding synthesis paragraphs[27]. For example, Figure 6 shows that the typical firing temperature is much lower when $FeC_2O_4$ is used as a precursor than when $Fe_2O_3$ is, whereas the firing temperature for CaO is comparable to that for $CaCO_3$. Utilizing the overlap between the distributions of temperatures for two precursors, a distance is defined as follows to describe the similarity between the two precursors.

$$d_{\text{temp}}(p_i^1, p_i^2) = 1 - \frac{\text{overlapping area of two temperature distribution}}{\text{total area of two temperature distribution}} \tag{9}$$

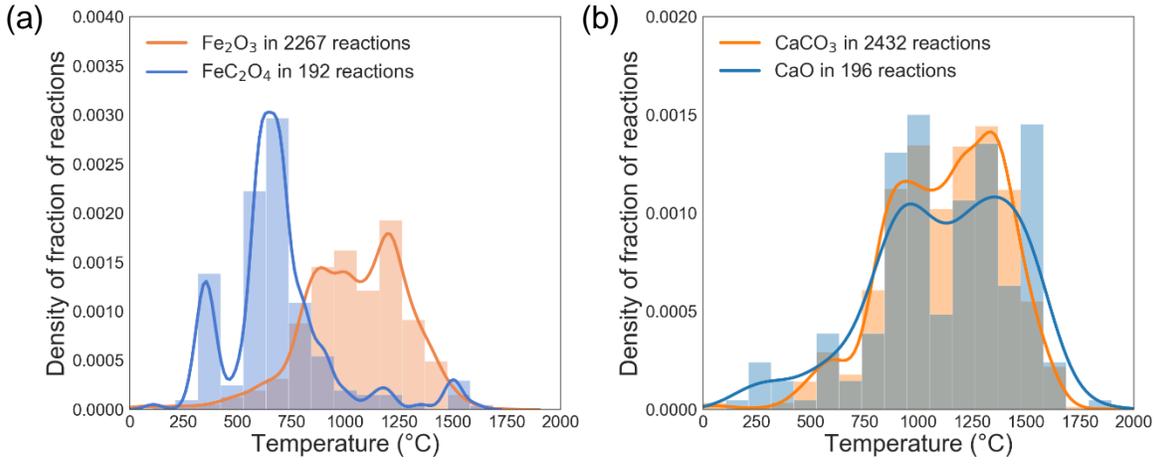

**Figure 6.** Highest firing temperature in the synthesis process for: (a) $Fe_2O_3$ and $FeC_2O_4$ and (b) $CaCO_3$ and CaO.

Both $d_{\text{sub}}$ in Eq. (8) and $d_{\text{temp}}$ in Eq. (9) satisfy the property that $0 \leq d_i \leq 1$. We utilized the Euclidean distance to define a multi-feature distance metric[61,62] to combine the two features together. The distance between a pair of precursors for the same element is defined as

$$D(p_i^1, p_i^2) = \sqrt{d_{\text{sub}}(p_i^1, p_i^2)^2 + d_{\text{temp}}(p_i^1, p_i^2)^2} \qquad (10)$$

The multi-feature aspect of this distance metric is general; it is straightforward to include additional features into this distance metric as new relevant features are considered. The current two representative features are selected because the substitution probability reflects the comparison of overall reactions in synthesis, and temperature is the most important parameter to activate these reactions. Finally, to visualize the similarity of precursors for the same element, we performed hierarchical clustering based on the pairwise distance $D(p_i^1, p_i^2)$ using Ward's minimum variance method[63]. The hierarchical clustering method iteratively identifies two nearest clusters and merges them until only one supercluster is left.

**4.2 Similarity of precursors.** Based on the distance defined in Eq. (10), precursors for the same elements were hierarchically clustered, and the similarities between them are displayed as dendrograms in Figure 7. The vertical axis represents the distance between two precursors or the distance between two clusters. In general, similar precursors will be drawn closer to each other on the horizontal axis.

Generally, the cluster with the smallest internal distance includes the common precursors, indicated using bold fonts in Figure 7. Simple binary fluorides and sulfides are far away from the common precursors and are typically used as source of F and S in target materials so that HF and $H_2S$ can be avoided. Metals are sometimes used as precursors directly; however, they are far away from the common precursors, indicating that metals and metal oxides tend to be used as precursors for different classes of materials. There is a trend that precursors are clustered following the order: oxide, carbonate, nitrate, and acetate, where the adjacent precursors are more similar (e.g., carbonate and oxide, or carbonate and nitrate), and the nonadjacent precursors are less similar (e.g. oxide and acetate) though there are variations to this for some elements. When the common precursor is a carbonate, the order may change to nitrate, carbonate, oxide, and acetate (e.g., Ba), where the carbonate and the nitrate are more similar than the carbonate and the oxide, but the carbonate still sits between the nitrate and the oxide. The similarity between different classes is possibly correlated with the different bonding strength between the cations and anions, which can be indicated by the order of melting points, namely, oxide > carbonate > nitrate/acetate.

However, there are also some observations that are not easy to immediately rationalize. For Li, it is the hydroxide rather than oxide or nitrate closest to the carbonate, whereas for Ca and Ba, the hydroxides are even absent, which means $Ca(OH)_2$ and $Ba(OH)_2$ are rarely used. This difference may originate from the methods used to prepare these precursors being different, resulting in different availabilities. One practical clue is that $Li_2O$ is more expensive than LiOH; $Li_2O$ (≥95% purity) is $378.00 for 100 g ($8.10/g of Li), while lithium hydroxide monohydrate (≥95% purity) is $181.00 for 2 kg ($0.54/g of Li) from the chemical supplier Strem Chemicals[64]. It is also observed that LiAc and $LiH_2PO_4$, as well as $FeC_2O_4$ and $FePO_4$, are clustered together, because they are frequently used to synthesize the extensively studied cathode material LiFePO$_4$, reflecting possible application bias in the data. In addition, oxides are similar to each other for variable valence elements, but the most similar precursor to the common oxide is not necessarily an oxide. For example, the oxides of Mn are clustered together, ranging from MnO$_2$ to MnO. However, the most similar precursor to MnO$_2$ is MnCO$_3$, as discussed in Section 3.4. Similarly, the nitrate Fe(NO3)$_3$ is more similar to Fe$_2$O$_3$ than the mixed-valence oxide Fe$_3$O$_4$ to Fe$_2$O$_3$. There are many factors in the selection of precursors, including both scientific reasons such as bonding, reactivity, and melting point, and anthropogenic reasons[65] such as literature success, convenience, applications, price, and human bias. The data in this work are a reflection of all those factors; it is not entirely clear how to deconvolute all these issues. An interesting scientific advance would be to identify the precursors that are chemically compelling while avoiding the implicit anthropogenic biases. This work provides a historical statistical analysis to serve as a baseline comparison.

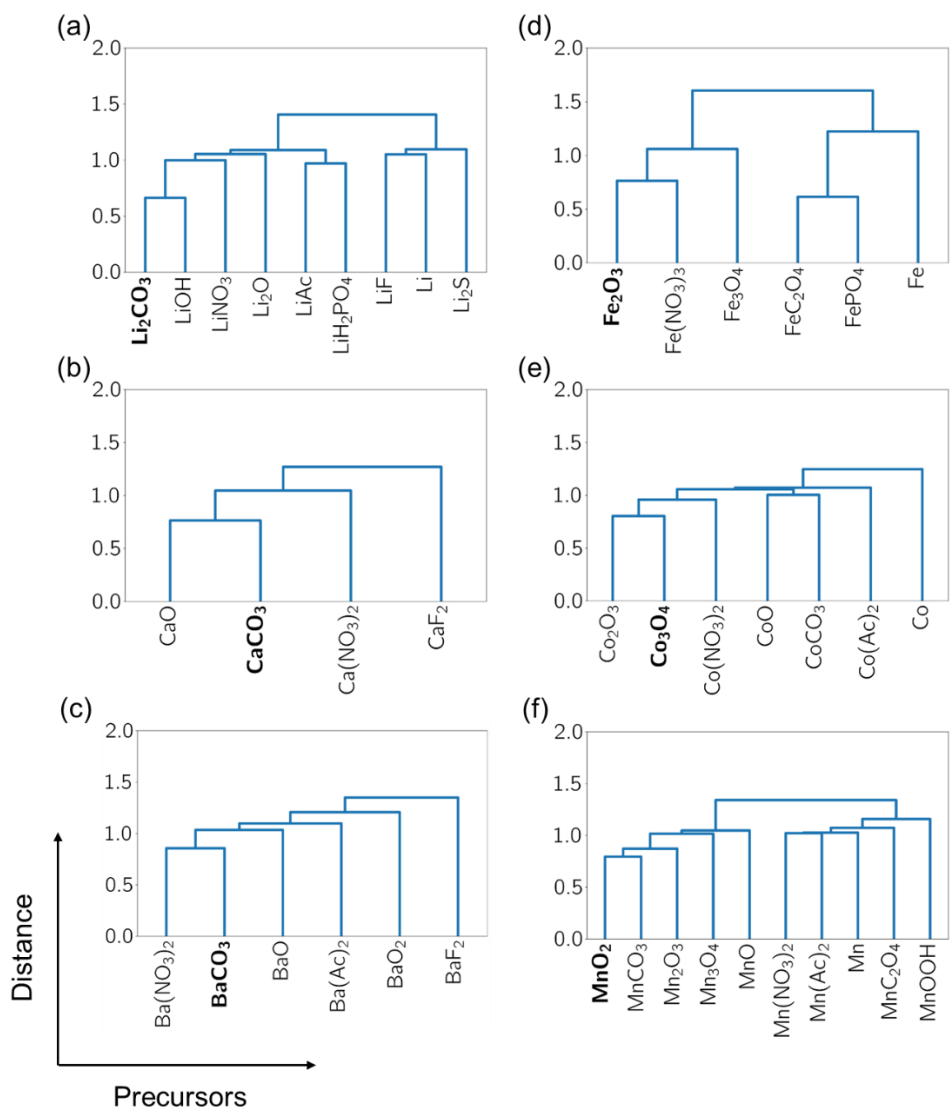

**Figure 7.** Clusters of precursors for (a) Li, (b) Ca, (c) Ba, (d) Fe, (e) Co, and (f) Mn by similarity. The common precursors are indicated using bold fonts.

The similarity could help guide the selection of precursors when researchers alter existing recipes by replacing precursors. For a starting experiment, it might be profitable to pick precursors similar to what has been tried before. On the other hand, when the synthesis is not going well, it is best to use a very different precursor in order to diversify the synthesis space. If there are many possible combinations of precursors, the quantitative value of the similarity could also serve as a reference to rank them. Currently, the creation of new recipes is in principle limited to targets already in our dataset. Therefore, it is also important in the future to develop similarity among targets. In that way, it would be possible to predict synthesis recipes for new target materials by evaluating the similarity with targets for which synthesis is known, a process that is very similar to the current literature-based approach for the synthesis of novel materials.

**5. CONCLUSIONS**

In this study, we proposed a two-step model based on Bi-LSTM to extract the precursors and targets in inorganic solid-state synthesis reactions as reported in 86,554 literature papers. The $F_1$ scores for the extraction of precursors and targets are 90.0% and 84.5%, respectively. Through comparison with a simple baseline model and showing how Bi-LSTM takes advantage of not only the written expression of words but also the surrounding context, we illustrated why the use of Bi-LSTM is suitable for our Chemical Named Entity Recognition (CNER) problem.

Using the extracted data, we conducted a meta-analysis on the similarities and differences between precursors. The statistics on the frequency to use different classes of precursors shows that each element usually has a common precursor to bring it into a target compound. A substitution model is used to quantify the probability of substituting one precursor with another while the target remains unchanged. By establishing distance metrics from the substitution model and the distribution of synthesis temperature, precursors for the same element were clustered to show the similarities between these precursors. This hierarchical clustering demonstrates that chemical domain knowledge of solid-state synthesis can be captured from text mining and provides a foundation for developing a predictive synthesis model.

**METHODS**

**Data preparation.** Borges[66] was used to scrape papers from websites of main publishers under agreements made with them. LimeSoup[67] was used to parse the papers from HTML content into plain text. Solid-state synthesis paragraphs were identified with the synthesis paragraph classification model[68] by Huo *et al.*[15]

**CNER and similarity analysis.** The SMR model was developed with Theano[69] and TensorFlow[70] based on the work by Lample *et al.*[29] An internal crowdsourcing website similar to Amazon Mechanical Turk[71] was built for data annotation. ChemDataExtractor[20] was used for text tokenization. Gensim[72] was used to train the Word2Vec[30,31] embeddings. The precursor substitution model was adapted from the ion substitution model developed by Hautier *et al.*[46] and Yang *et al.*[47] as in pymatgen[60]. All coding was with Python 3[73]. More details of the methods are introduced in each section.

## ASSOCIATED CONTENT

**Supporting Information.** More discussion on benefits of the two-step model, comparison with BERT, and utilization of synthesis time. This material is available free of change via the Internet at http://pubs.acs.org.

## AUTHOR INFORMATION

**Corresponding Author**

**Gerbrand Ceder –** Department of Materials Science and Engineering, UC Berkeley, Berkeley, CA, 94720, USA; Materials Sciences Division, Lawrence Berkeley National Laboratory, Berkeley, CA, 94720, USA; E-mail: gceder@berkeley.edu

## CODE AVAILABILITY

The code of the Synthesis Materials Recognizer (SMR) model is publicly available at the GitHub repository https://github.com/CederGroupHub/text-mined-synthesis_public.

## ACKNOWLEDGMENT

Funding to support this work was provided by the Energy & Biosciences Institute through the EBI-Shell program (Award No PT74140 and PT78473), the Assistant Secretary of Energy Efficiency and Renewable Energy, Vehicle Technologies Office, U.S. Department of Energy under Contract No. DE-AC02-05CH11231, the Office of Naval Research (ONR) Award #N00014-16-1-2432, the National Science Foundation under Grant Number 1922311, 1922372, and 1922090, and the U.S. Department of Energy, Office of Science, Office of Basic Energy Sciences, Materials Sciences and Engineering Division under Contract No. DE-AC02-05-CH11231 (D2S2 program KCD2S2). This work used Savio computational cluster resource provided by the Berkeley Research Computing program at the University of California, Berkeley (supported by the UC Berkeley Chancellor, Vice Chancellor for Research, and Chief Information Officer), and the Extreme Science and Engineering Discovery Environment (XSEDE), which is supported by National Science Foundation grant number ACI-1548562. Specifically, it used the Bridges system, which is supported by NSF award number ACI-1445606, at the Pittsburgh Supercomputing Center (PSC). We thank Anna Sackmann (Science Data and Engineering Librarian at UC Berkeley) for helping us to obtain Text and Data Mining agreements with the specified publishers. We also thank Prof. Elsa Olivetti, Edward Kim, Alexander Van Grootel, and Zach Jensen for valuable collaborations and help with content acquisition and HTML/XML markups parser development, and Chris Bartel, Zheren Wang, Amalie Trewartha, Nicolas Mingione, Haegyeom Kim, Guobo Zeng, Huiwen Ji, Indranil Rudra, Padmini Rajagopalan, Kaustubh Kaluskar, and Lalit Gupta for valuable discussions. Finally, we thank all the Ceder group members for their help with data annotation and manual checks of the data.

# TABLE OF CONTENTS (TOC) GRAPHIC

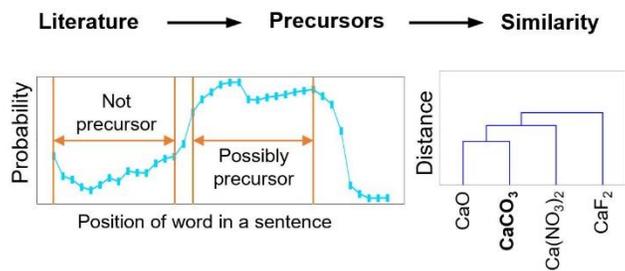